\documentclass[sigconf]{acmart}

\acmConference[ICPC 2022]{The 30th International Conference on Program Comprehension}{May 21–22, 2022}{Pittsburgh, PA, USA}

\AtBeginDocument{%
	\providecommand\BibTeX{{%
			\normalfont B\kern-0.5em{\scshape i\kern-0.25em b}\kern-0.8em\TeX}}}

\setcopyright{acmcopyright}
\copyrightyear{2022}
\acmYear{2022}
\acmDOI{10.1145/3524610.3527907}

\acmConference[ICPC '22]{30th International Conference on Program Comprehension}{May 16--17, 2022}{Virtual Event, USA}
\acmBooktitle{30th International Conference on Program Comprehension (ICPC '22), May 16--17, 2022, Virtual Event, USA}
\acmPrice{15.00}
\acmISBN{978-1-4503-9298-3/22/05}
\usepackage{multirow, tabularx}
\usepackage{subfigure}
\begin{document}
	
	\bibliographystyle{unsrt}
	
	%%
	%% The "title" command has an optional parameter,
	%% allowing the author to define a "short title" to be used in page headers.
	\title{M2TS: Multi-Scale Multi-Modal Approach Based on Transformer for Source Code Summarization}

	\author{Yuexiu Gao}
	\affiliation{%
		\institution{Shandong Normal University}
		\city{Jinan}
		\country{China}}
	% \email{yxgao8@163.com}  
	
	\author{Chen Lyu}
	% \orcid{0000-0002-5044-1459}
	% \authornotemark[1]
	\authornote{Corresponding author. Email: lvchen@sdnu.edu.cn}
	\affiliation{%
		\institution{Shandong Normal University}
		%   \streetaddress{P.O. Box 1212}
		%   \city{Dublin}
		%   \state{Ohio}
		%   \country{USA}
		\city{Jinan}
		\country{China}
		%   \postcode{43017-6221}
	}
	% \email{lvchen@sdnu.edu.cn}
	% %%
	% %% By default, the full list of authors will be used in the page
	% %% headers. Often, this list is too long, and will overlap
	% %% other information printed in the page headers. This command allows
	% %% the author to define a more concise list
	% %% of authors' names for this purpose.
	\renewcommand{\shortauthors}{Gao, et al.}
	
	%%
	%% The abstract is a short summary of the work to be presented in the
	%% article.
	\begin{abstract}
		Source code summarization aims to generate natural language descriptions of code snippets. Many existing studies learn the syntactic and semantic knowledge of code snippets from their token sequences and Abstract Syntax Trees (ASTs). They use the learned code representations as input to code summarization models, which can accordingly generate summaries describing source code. Traditional models traverse ASTs as sequences or split ASTs into paths as input. However, the former loses the structural properties of ASTs, and the latter destroys the overall structure of ASTs. Therefore, comprehensively capturing the structural features of ASTs in learning code representations for source code summarization remains a challenging problem to be solved. In this paper, we propose M2TS, a \textbf{M}ulti-scale \textbf{M}ulti-modal approach based on \textbf{T}ransformer for source code \textbf{S}ummarization. M2TS uses a multi-scale AST feature extraction method, which can extract the structures of ASTs more completely and accurately at multiple local and global levels. To complement missing semantic information in ASTs, we also obtain code token features, and further combine them with the extracted AST features using a cross modality fusion method that not only fuses the syntactic and contextual semantic information of source code, but also highlights the key features of each modality. We conduct experiments on two Java and one Python datasets, and the experimental results demonstrate that M2TS outperforms current state-of-the-art methods. We release our code at \url{https://github.com/TranSMS/M2TS}.
	\end{abstract}
	
	%%
	%% The code below is generated by the tool at http://dl.acm.org/ccs.cfm.
	%% Please copy and paste the code instead of the example below.
	%%
	\begin{CCSXML}
		<ccs2012>
		<concept>
		<concept_id>10010520.10010553.10010562</concept_id>
		<concept_desc>Software and its engineering~Software maintenance tools</concept_desc>
		<concept_significance>500</concept_significance>
		</concept>
		<concept>
		<concept_id>10010520.10010575.10010755</concept_id>
		<concept_desc>Computer systems organization~Redundancy</concept_desc>
		<concept_significance>300</concept_significance>
		</concept>
		<concept>
		<concept_id>10010520.10010553.10010554</concept_id>
		<concept_desc>Computer systems organization~Robotics</concept_desc>
		<concept_significance>100</concept_significance>
		</concept>
		<concept>
		<concept_id>10003033.10003083.10003095</concept_id>
		<concept_desc>Networks~Network reliability</concept_desc>
		<concept_significance>100</concept_significance>
		</concept>
		</ccs2012>
	\end{CCSXML}
	
	\ccsdesc[500]{Software and its engineering~Software maintenance tools}
	
	% \ccsdesc[300]{Computer systems organization~Redundancy}
	% \ccsdesc{Computer systems organization~Robotics}
	% \ccsdesc[100]{Networks~Network reliability}
	
	%%
	%% Keywords. The author(s) should pick words that accurately describe
	%% the work being presented. Separate the keywords with commas.
	\keywords{Source code summarization, Transformer, Neural network, Deep learning}
	
	%% A "teaser" image appears between the author and affiliation
	%% information and the body of the document, and typically spans the
	%% page.
	% \begin{teaserfigure}
	%   \includegraphics[width=\textwidth]{sampleteaser}
	%   \caption{Seattle Mariners at Spring Training, 2010.}
	%   \Description{Enjoying the baseball game from the third-base
	%   seats. Ichiro Suzuki preparing to bat.}
	%   \label{fig:teaser}
	% \end{teaserfigure}
	
	%%
	%% This command processes the author and affiliation and title
	%% information and builds the first part of the formatted document.
	\maketitle
	
	\section{Introduction}
	During software development and maintenance, more than half of the time is spent on program understanding and related tasks \cite{corbi1989program}. In most of these tasks, developers resort to looking at comments to comprehend the meaning of source code \cite{xia2017measuring}. However, the writing of comments is often overlooked in software development, resulting in poor quality comments being obtained by developers \cite{ko2004six, ko2006exploratory}. Source code summarization technology is the generation of brief natural language descriptions for source code \cite{latoza2006maintaining}. Its emergence not only frees developers from handwritten comments but also increases the efficiency and reduces the cost of software development \cite{eddy2013evaluating, mcburney2015automatic, singer2010examination, roscheisen1998stanford}.
	
	The earliest approaches to source code summarization are traditional, which rely mainly on templates \cite{wong2015clocom, wong2013autocomment, rodeghero2015eye, rodeghero2014improving}. Sridhara et al. \cite{sridhara2010towards} applied a method for creating artificial templates where they used the Software Word Usage Model (SWUM) to generate comments for Java methods. Haiduc et al. \cite{haiduc2010supporting, haiduc2010use} used an Information Retrieval (IR) approach to generate summaries, they applied a Vector Space Model (VSM) to search summaries from similar code fragments. Although these methods have achieved specific results, they are overly dependent on naming normality and will not work if no similar code fragments are given in the codebase.
	
	With the rapid development of deep learning, extensive research on source code summarization has emerged \cite{hu2018summarizing, guo2020multi, wei2020retrieve, xie2021exploiting}. Lyer et al. \cite{iyer2016summarizing} used an attention-based sequence to sequence model to generate summaries for source code. Later, in order to learn the structural features of source code, some studies have used API \cite{hu2018summarizing} sequences or ASTs \cite{wan2018improving} as input, Figure \ref{fig:R1} shows a code snippet and its AST. Hu et al. \cite{hu2018deep, hu2020deep} and LeClair et al. \cite{leclair2019neural} fed AST sequences obtained by traversal into the encoder-decoder model to generate summaries. Alon et al. \cite{alon2018code2seq} extracted paths from ASTs to help generate comments. Although these studies are carried out on ASTs, they have destroyed the structural properties and integrity of ASTs. Both the encoder and decoder used in the above studies were RNNs (LSTM or GRU) \cite{hochreiter1997long, cho2014learning}. To address the inability of RNNs to capture the long dependencies arising from using long sequences as input, Vaswani et al. \cite{vaswani2017attention} proposed a Transformer model, which is effective on neural machine translation tasks. Ahmad et al. \cite{ahmad2020transformer} improved on the ordinary Transformer and obtained excellent results using only source code as input of the model. Most of the above works \cite{hu2020deep, leclair2019neural, leclair2020improved, zhang2020retrieval} have shown that using ASTs and source code as input can improve the quality of generated summaries, so we also use two modalities for study in this paper.
	
	Two factors that influence the quality of summary generation can be identified from the above studies:
	
	\textbf{AST structural feature extraction.} The treatment of ASTs in the previous studies mainly includes traversing them into sequences or using AST paths. Hu et al. \cite{hu2018deep, hu2020deep} used AST sequences obtained by a Structure-Based Traversal (SBT) method as input, which only extracts the AST structure globally without considering the relationships between nodes, losing the structural properties of ASTs \cite{yang2021multi}. In addition, LeClair et al. \cite{leclair2020improved} used GNN to embed the AST graph, but the embedding of ASTs using 2-Hop GNN was unable to extract the overall structure of ASTs. So the first problem we need to solve is how to extract the structural features of ASTs more comprehensively. 
	
	% \vspace{-0.1cm}
	\textbf{Multi-modal feature fusion.} There are numerous studies in the source code summarization that use multi-modal features of source code for modeling, where the multi-modal refers to multiple different representations of source code. For example, Hu et al. \cite{hu2020deep}, Zhang et al. \cite{zhang2019novel}, LeClair et al. \cite{leclair2020improved} used the AST and source code for their research, and the semantic and syntactic information of code can be learned using these two modalities. In the existing multi-modal fusion methods, most of them use the traditional attention mechanism to sum up different modality features \cite{hu2020deep, leclair2019neural}. However, this method allocates attention only within modalities while ignoring the impact of inter-modal importance on the generated summaries, failing to highlight critical information in the source code. Therefore, how to better fuse multiple modal features is the second issue we need to address.
	
	For the first problem, we use multiple scales of ASTs (i.e., the different power matrices obtained from the AST adjacency matrix) combined with the Graph Convolutional Neural Network (GCN) for feature extraction. Since the adjacency matrix only represents the relationship between neighboring nodes while not obtaining the information of non-neighboring nodes, the multi-scale method can ensure the comprehensive AST structural features extracted at multiple local and global levels. For the second problem, we use a new cross modality feature fusion method that not only fuses the syntactic and contextual semantic knowledge of source code, but also highlights key features that help generate summaries.
	
	In this paper, synthesizing the above issues, we propose a multi-scale multi-modal approach based on Transformer for source code summarization. First, we propose a multi-scale AST feature extraction method that uses GCN to embed ASTs at multiple scales. Then the embedded AST and code tokens vectors are encoded respectively, and these encoded features are fused using a new cross modality feature fusion method. Finally, the fused features are fed into the decoder with the encoded code tokens features to generate summaries.
	
	We conduct extensive experiments on the three datasets, and the experimental results demonstrate that M2TS obtains better scores than current state-of-the-art methods on BLEU, METEOR, ROUGE\_L and CIDER evaluation metrics. For example, our model obtains a 57.87\% ROUGE\_L score on the Java dataset used by Hu et al. \cite{hu2018summarizing}, which is 3.8\% higher than the current state-of-the-art method (SG-Trans), and 33.84\% BLEU-4 score on the Python dataset used by Barone et al. \cite{leclair2019recommendations}, which is 5.1\% higher than SG-Trans. In addition, we also conduct ablation experiments on M2TS, and the results show that our proposed multi-scale and multi-modal method is essential for improving the model performance.
	
	Our main contributions are as follows:
	
	\begin{itemize}
		\item We propose a Transformer-based multi-scale multi-modal approach for source code summarization (M2TS) that comprehensively extracts semantic and syntactic structural features of source code, resulting in higher quality generated summaries.
		\item We present a multi-scale AST feature extraction method, which can extract the structure of ASTs more completely and accurately from both multiple local and global levels.
		\item We design a new cross modality feature fusion method that not only highlights the structural features of source code but also learns the contextual relevance between code tokens.
		\item We conduct extensive experiments including quantitative and qualitative comparisons that show the effectiveness of M2TS when compared with other state-of-the-art methods. 
	\end{itemize}
	
	% \vspace{-0.595cm}
	\section{Background and Motivation}
	
	\subsection{The Transformer Structure}
	The Transformer is a relatively popular model in recent years \cite{vaswani2017attention}, and it has been widely used in fields such as natural language processing and computer vision. The following are the details of several significant structures used in the Transformer.
	% Unlike the structures such as LSTM and GRU used in previous studies \cite{hu2018deep, leclair2019neural, leclair2020improved}, the encoder and decoder in the Transformer are composed of multi-head attention mechanisms and fully connected networks.
	
	\subsubsection{Multi-head Attention.} The structure is the main component in the Transformer model, and it solves the long dependency problem of sequences. Multi-head attention is the set of multiple self-attentions. In self-attention, the input vector is respectively multiplied with the weights $ W^q, W^k, W^v$ to obtain three new vectors $Q, K$ and $V$. While in multi-head attention, multiple $Q, K$ and $V$ are obtained by multiplying the input vector with several different weights. The following is the calculation formula of multi-head attention:
	\begin{equation}
		Q_{i} = XW^q_{i},\quad K_{i} = XW^k_{i} ,\quad V_{i} = XW^v_{i},
	\end{equation}
	\begin{equation}
		head_{i} = Attention(Q_i, K_i, V_i) = softmax(\frac{Q_iK^T_{i}}{\sqrt{d_k}})V_i,
	\end{equation}
	\begin{equation}
		MultiHead(Q, K, V) = Concat(head_1, \dots, head_n)W^O,
	\end{equation}
	where $X$ is the input vector, $i$ denotes the $i\_th$ head, and $n$ denotes the total number of heads \cite{vaswani2017attention}.
	
	\subsubsection{Position Encoding.} Since the Transformer does not use RNNs, it cannot use the order information of words. Therefore, positional encoding is used in the Transformer to preserve the position of the word in the sequence. We use $PE$ to denote position encoding. The dimension of $PE$ is kept consistent with the input vector. The position encoding can be obtained by training or formula. The Transformer use the following formula to get the position encoding:
	\begin{equation}
		PE_{(pos,2i)} = \sin (pos/10000^{2i/d}),
	\end{equation}
	\begin{equation}
		PE_{(pos,2i+1)} = \cos (pos/10000^{2i/d}),
	\end{equation}
	where $pos$ denotes the position of the word in the sentence, $d$ denotes the dimension of $PE$, $2i$ denotes the even dimension, and $2i+1$ denotes the odd dimension \cite{vaswani2017attention}.
	
	\begin{figure}[htbp]
		\centering
		\subfigure[A Java method and its AST]{
			\includegraphics[width=0.89\linewidth]{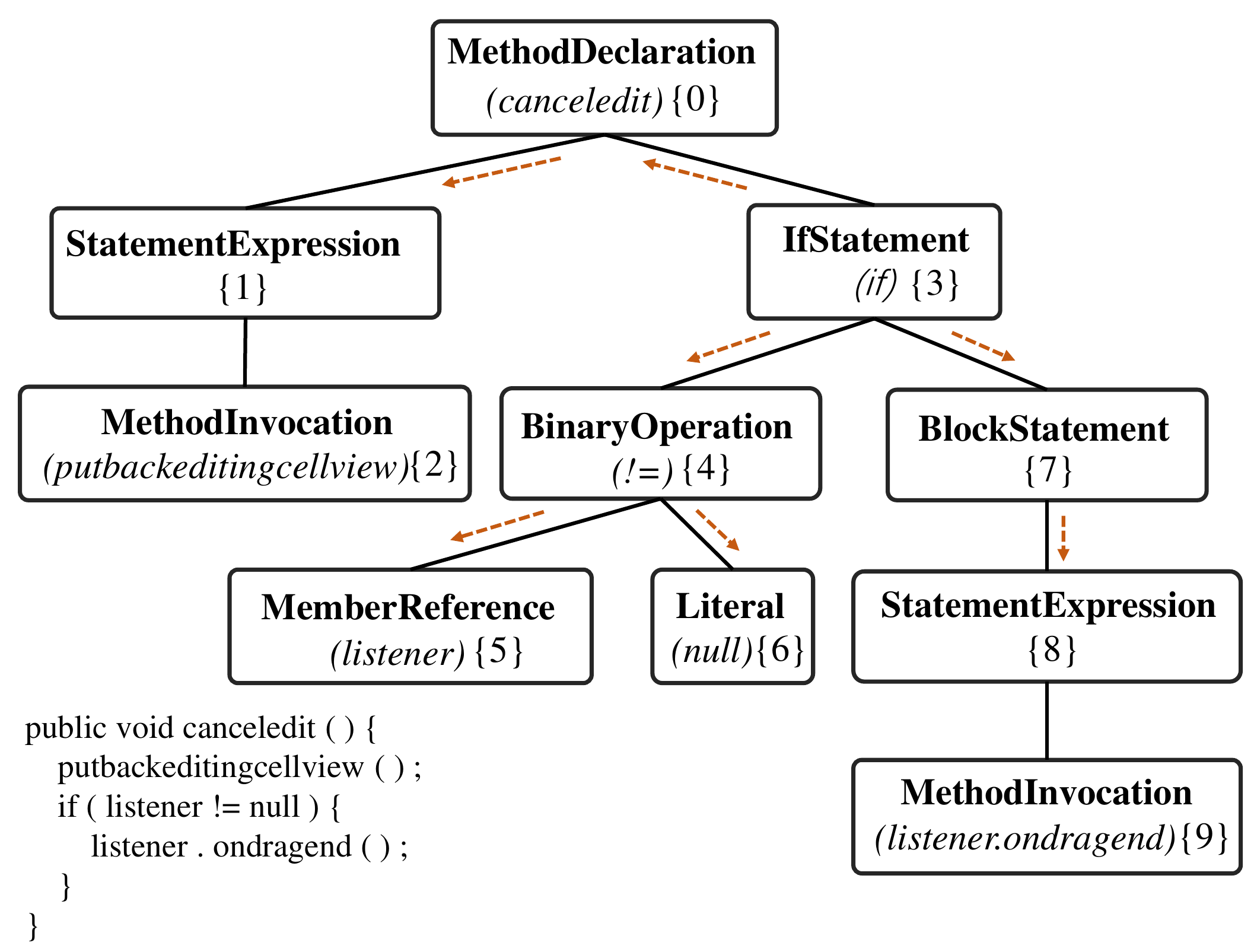}
			\label{fig:R1}
		}
		\subfigure[The adjacency matrix and its quadratic matrix]{
			\includegraphics[width=0.89\linewidth]{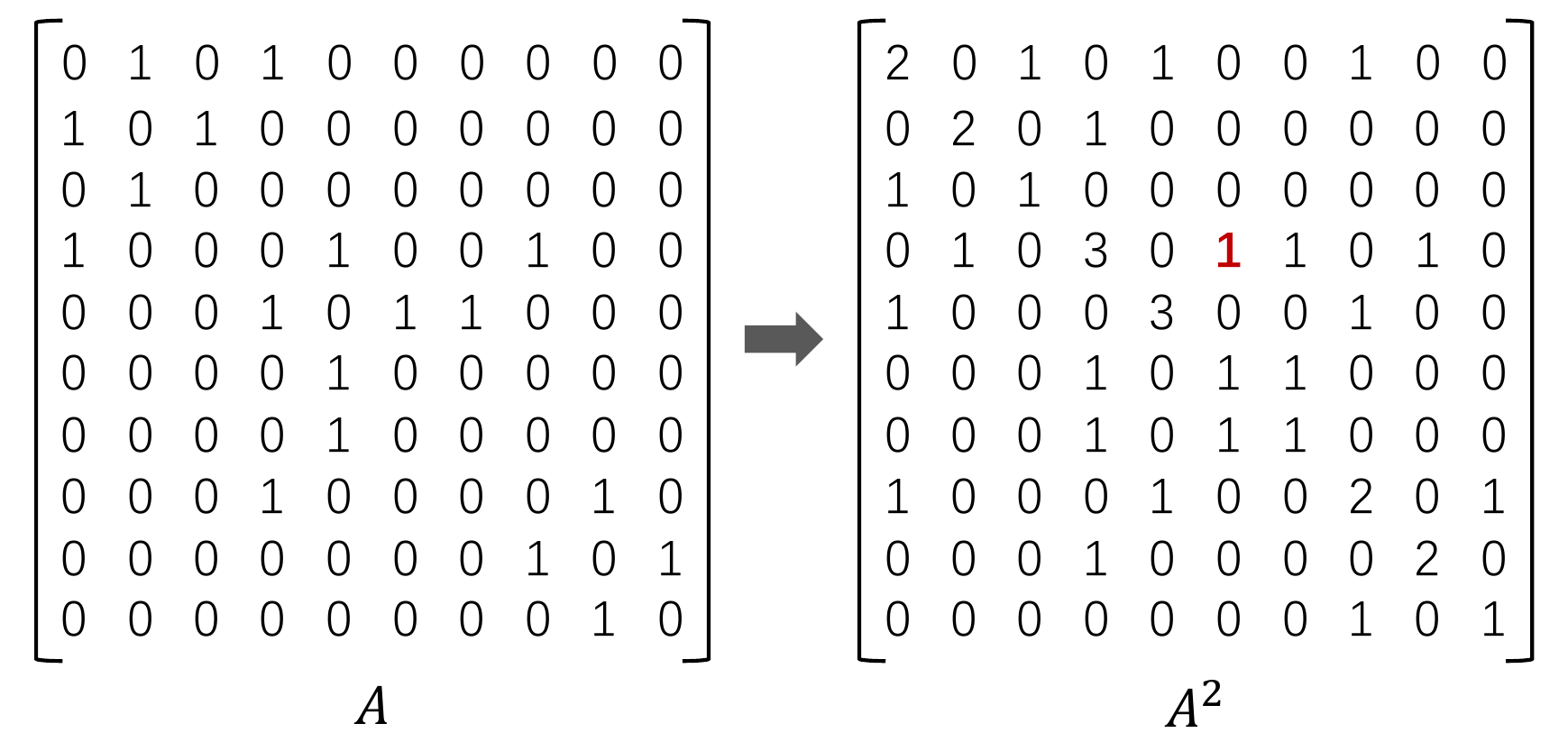}
			\label{fig:R2}
		}
		\caption{A motivating example. \subref{fig:R1} is a Java method and its AST, the type of nodes is in boldface and its corresponding value of nodes is in italics, the number represents the ID of AST nodes; \subref{fig:R2} represents the adjacency matrix corresponding to AST in \subref{fig:R1} and its quadratic matrix.}
		\label{fig:figure1}
	\end{figure}
	
	\subsection{Motivating Example}
	To explain that our proposed multi-scale method is effective, we use a specific example for illustration in this subsection. We have learned that code structures are crucial for generating summaries by analyzing previous studies \cite{alon2018code2seq, hu2018deep, leclair2020improved, xu2018graph2seq}. In the research on the representation of ASTs syntax structural features, we summarize three common methods:
	
	\begin{enumerate}
		\item Traversal of ASTs; 
		\item Using ASTs paths;
		\item Embedding ASTs using Graph Neural Networks.
	\end{enumerate}
	
	Although the above three methods have achieved certain results, they also have corresponding shortcomings. In the following, we will use Figure \ref{fig:figure1} to illustrate the weaknesses of these methods and demonstrate the advantages of our new method. For convenience, we use the node ID to represent the entire node in the following exposition.
	
	The first is to traverse ASTs to sequences. The AST sequence that can be obtained using traditional traversal methods (e.g., prior-order traversal) for Figure \ref{fig:R1} is 0 1 2 3 4 5 6 7 8 9. However, this traversal method has two significant problems, one is that the two nodes with a common parent are far apart by traversal (for example, children nodes 4 and 7 of node 3), and the other is that the sequence may not be restored to a unique AST. To solve the above problems, Hu et al. \cite{hu2018deep} proposed the SBT method, the sequence of Figure \ref{fig:R1} was obtained by using this method as (0 (1 (2) 2) 1 (3 (4 (5) 5 (6) 6) 4 (7 (8 (9) 9) 8) 7) 3) 0, which can restore a unique AST. However, compared with the traditional traversal methods, the sequences obtained by the SBT method are longer, which inevitably leads to the long dependence problem when encoding. In addition, this way of using AST sequences as input of model destroys the structural properties of ASTs.
	
	In the methods of using ASTs paths, each path of ASTs is usually obtained and then encoded \cite{alon2018code2seq}. We can get several paths of Figure \ref{fig:R1} as: 0→1→2, 0→3→4→5, 0→3→4→6, 0→3→7→8→9. The AST paths are encoded in such a way that they preserve the hierarchical relationships of nodes, but this method also has certain limitations. For example, the children nodes 4 and 5 of node 2, which are in two pathways respectively, result in the correlation of the two nodes being lost when encoded. Therefore, this method breaks the integrity of the AST structure to some extent.
	
	The last is embedding ASTs using GNNs \cite{yang2021multi, gao2021code}, and here we take GCN as an example. As the number of GCN layers increases, each node can aggregate a more extensive range of information from its neighbors, thus focusing on a broader range of local features. For example, two layers of GCN are used for Figure \ref{fig:R1} AST, and node 0 can aggregate not only the information of nodes 1 and 3 but also the information of nodes 2, 4 and 7 indirectly. Although GCN can extract the local structure of ASTs \cite{peng2021integrating}, it ignores the global information of ASTs.
	
	To address the problems in the above studies, we propose a multi-scale AST feature extraction method. The multi-scale refers to the different power matrices of the AST adjacency matrix by dot product. These power matrices have the special property that the $m\_th$ power matrix represents how many schemes there are between nodes arriving through $m$ edges. For example, the element 1 in the red font of $A^2$ in Figure \ref{fig:R2} indicates that one scheme is reached between nodes 3 and 5 through two edges. Based on the obtained power matrices, it can be seen from the AST that a certain node can establish a specific local structure with other nodes. As shown in Figure \ref{fig:figure1}, the $A^2$ in Figure \ref{fig:R2} shows that reaching node 3 through two edges are nodes 1, 5, 6, 8, and node itself. It can be seen from the Figure \ref{fig:R1} AST that these edges can represent the syntactic structure of the node "\textit{IfStatement}" (the orange dashed line indicates the reachable paths), and this local structure can be extracted by embedding the $A^2$ matrix using GCN. Unlike general GCN which only uses the adjacency matrix to embed the local features of ASTs. In contrast, our method can represent multiple different local structures of nodes by using multiple scales of ASTs, and then using the weighted summation method introduced in Section 3.1 also can extract the global structural features of ASTs. 
	
	% \vspace{-0.895cm}
	\section{our Approach}
	\begin{figure*}[htbp]
		\centering
		\includegraphics[width=0.81\textwidth]{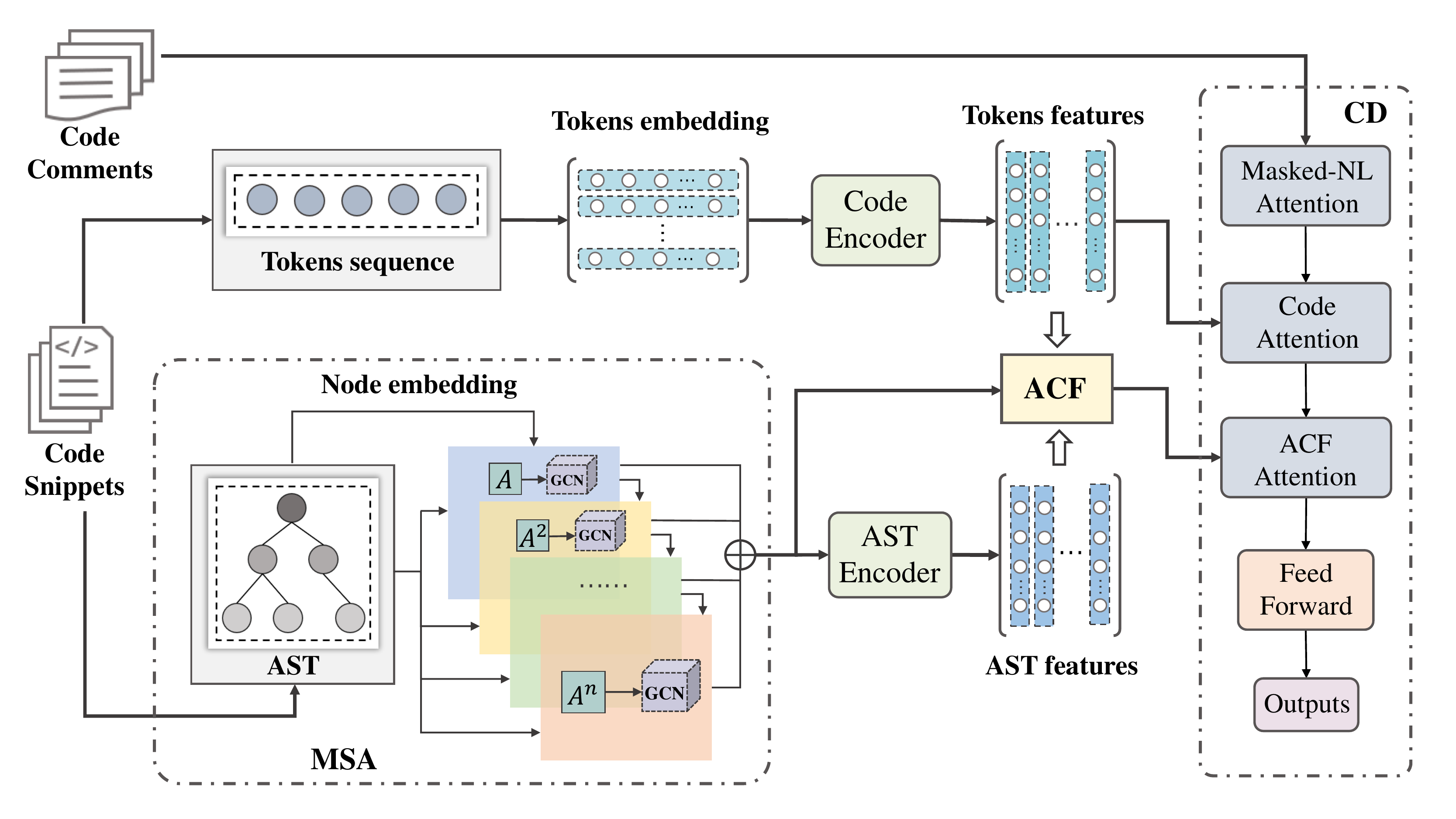}
		\caption{The overall framework of our approach}
		\label{fig:figure2}
	\end{figure*}
	Our proposed approach M2TS mainly consists of a \textbf{M}ulti-\textbf{S}cale \textbf{A}ST feature extraction module (MSA), a cross modality feature fusion module (ACF, i.e., \textbf{A}ST and \textbf{C}ode \textbf{F}usion), and a \textbf{C}ombined \textbf{D}ecoder (CD). Among them, the MSA module embedding ASTs at multiple scales can extract structural features of the code more completely and accurately. The ACF module fuses features of two modalities after encoding, which not only highlights the structural information of the code but also learns the contextual semantic relatedness between code tokens. Finally, the fused features and the encoded code tokens features are fed into the combined decoder to generate summaries. The decoder can be used for information supplementation. The overall framework of M2TS is illustrated in Figure \ref{fig:figure2}.
	
	\subsection{Multi-Scale AST Feature Extraction}
	In this paper, we use GCN to embed the multiple scales AST. We have learned that the best results can be achieved by using a two-layer GCN for the node classification problem \cite{kipf2016semi}, so in this paper, we also use a two-layer GCN for the AST embedding. Unlike the general multi-layer GCN, we add the output of the upper layer graph convolution to the output of the current layer using a residual connection. Each graph convolution layer is calculated using the following formula:
	\begin{equation}
		H^{(l+1)} = \sigma(\hat{A}H^{(l)}W^{(l)}) + H^{(l)},
	\end{equation}
	where $l$ denotes the number of layers of the graph convolution, $\hat{A}$ denotes the normalized adjacency matrix, $H^{(l)}$ denotes the output of the previous layer of the graph convolution, $W^{(l)}$ denotes a weight matrix, and $\sigma$ is a nonlinear activation function \cite{kipf2016semi}.
	
	When inputting the first layer of graph convolution, we use the BERT pre-training \cite{devlin2018bert} to embed the "\textit{type}" and "\textit{value}" of nodes as the initial feature vector $H^{(0)}$. Here the vectors obtained by pre-training are fixed and do not participate in the whole optimization process of the model.
	
	From Section 2, according to the particular property of the power matrices, we regard the adjacency matrix $A$ and the power matrices $ A^2, \dots, A^n$ obtained by dot product as the first, second, and $n\_th$ scales of the AST, respectively. First, we input the first scale representation $A$ of the AST combined with $H^{(0)}$ into the two-layer GCN to obtain $Z_1$. After that, the second scale representation $A^2$ combined with the previous scale output $Z_1$ is input into the two-layer GCN to obtain $Z_2$. By analogy, the output $Z_{n-1}$ of the ${n-1}\_th$ scale representation $A^{n-1}$ is combined with the $n\_th$ scale representation $A^n$ into the same layer GCN to obtain the output $Z_n$. Finally, we weight the output of each scale and sum them up to obtain the final AST embedding representation $Z$. For example, when the number of scales $n$ is set to three, the AST embedding vector is calculated as:
	\begin{equation}
		Z = \alpha Z_1 + \beta Z_2 + (1- \alpha - \beta)Z_3,
	\end{equation}
	where the values of $\alpha$ and $\beta $ will be determined in the experiments. We use multiple scales AST representation in the MSA module, which extracts AST structural features at multiple local and global levels.
	%\vspace{-0.4cm}
	\subsection{Cross Modality Feature Fusion}
	To obtain more code features, we use the source code and AST modalities as input to M2TS. In order to fuse the features of two modalities, in this paper we use a new cross modality feature fusion method. The two modalities features need to be encoded before fusion.
	\begin{figure}[htbp]
		\centering
		\includegraphics[width=0.47\linewidth]{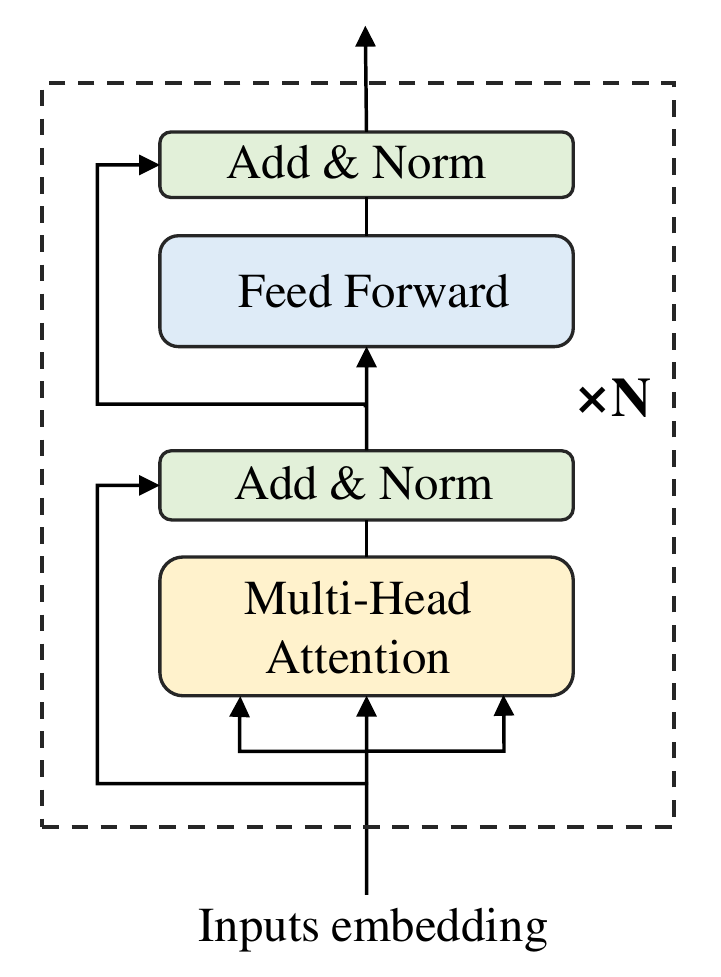}
		\caption{Basic structure of the two encoders. $N$ indicates the number of layers}
		\label{fig:figure3}
	\end{figure}
	We input the embedded vector $Z$ obtained from the MSA module into the AST encoder for encoding. Here we use a multi-head attention mechanism in the Transformer and a two-layer fully connected network (Feed Forward block in Figure \ref{fig:figure2}) as the primary constituent structure of the AST encoder. Before inputting into the encoder, we need to obtain the position encoding of the embedded feature of each node to preserve the position information of nodes. After encoding, we get the feature $Z'$. For the code tokens, we encode them using a code encoder, which has a similar structure to the AST encoder. Before entering the code encoder, we use word embedding to represent the code tokens. Then adding the word vectors and the position encoding of the tokens input to the code encoder to get the output $M$. It is worth noting that although the two encoders have similar structures, they are aimed at inputs of different modality features, so their parameters are optimized separately during the training process. The basic structure of these two encoders is shown in Figure \ref{fig:figure3}.
	
	Given the encoded AST feature $ Z' $ and the code tokens feature $M$, we fuse them into a unified representation using the ACF module, which is similar to the self-attention mechanism in the Transformer. We regard the AST feature $ Z'$ as the query and the code tokens feature $M$ as the key and value. Specifically, we first represent $ Z'$ and $M$ into $Q_f, K_f$ and $V_f $ by $1 \times 1$ convolution, respectively \cite{wang2021m2tr}:
	%\vspace{-0.3cm}
	\begin{equation}
		Q_f = Conv_{1 \times 1}( Z'), K_f = Conv_{1 \times 1}(M), V_f = Conv_{1 \times 1}(M),
	\end{equation}
	where $Q_f, K_f, V_f $ retain the same size as the original input and then calculate the fused feature by the following formula:
	\begin{equation}
		F = softmax(\frac{Q_fK^T_f}{\sqrt{d_k}})V_f,
	\end{equation}
	where $ d_k $ we set to 64.
	
\begin{figure}[htbp]
		\centering
		\includegraphics[width=0.85\linewidth]{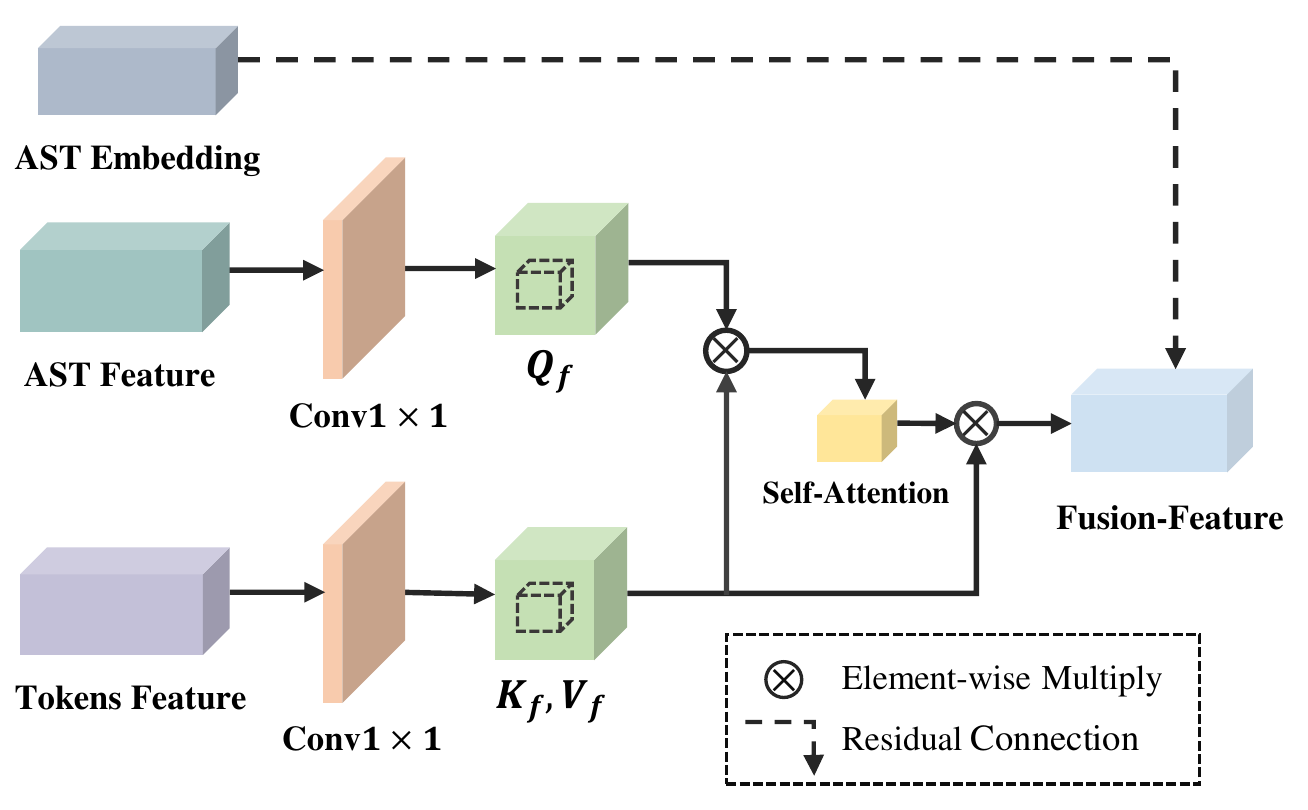}
		\caption{ACF module fusion process}
		\label{fig:figure4}
	\end{figure}

	Finally, we employ a residual connection to add the embedded AST representation $Z$ and the fused feature $F$ to obtain the final output $F'$. The detailed process of the ACF module is shown in Figure \ref{fig:figure4}.
	
	\subsection{Combined Decoder}
	We input the encoded code tokens and fused features into the combined decoder to generate summaries. In the training phase, the CD module consists of three attention blocks and a two-layer fully connected network: Masked-NL Attention, Code Attention, ACF Attention and Feed-Forward.
	 As with the two encoders used, a normalization is added after each block output in the decoder. In the following, we describe in detail the input and output of information in the CD module \cite{vaswani2017attention, sun2020treegen}.

	Generally, we input the natural language descriptions after word embedding and positional encoding into the Masked-NL Attention block, where the inputs $Q, K,$ and $V$ are all Natural Language (NL) embedding. This representation is a way of computation in the self-attention mechanism. After that, using the output $S_1$ of Masked-NL Attention block as $Q$ and the encoded code tokens feature $M$ as $K$ and $V$ input to the Code Attention block to get the output $S_2$. Subsequently, we input the output $S_2$ of the Code Attention block as $Q$ and the output $F'$ obtained from the ACF module as $K, V$ into the ACF Attention block to get the output $S$.
	
	The probability of generating summaries is then calculated using a softmax activation function after passing through a two-layer fully connected network. Using this decoder allows for information supplementation. In the CD module, we aim to minimize the cross-entropy loss function and update the parameters in the model using optimization algorithms such as gradient descendent. The following is the loss function:
	
	\begin{equation}
		L_{(loss)} = -\frac{1}{N} \sum\limits_{i=1}^N \sum\limits_{j=1}^n \log p(y^{(j)}_i), 
	\end{equation}
	where $N$ is the total number of training data, $n$ is the length of each target summary, $ y^{(j)}_i $ is the $j\_th$ word in the $i\_th$ sentence, and $ p(y^{(j)}_i)$ denotes the probability of generating the $j\_th$ word.

	% \begin{table}[htbp]
	% 	\caption{Comparison of setting different heads on the JAH dataset}
	% 	\label{tab:table4}
	% 	\resizebox{1\columnwidth}{!}{
	% 	\begin{tabular}{lccccc}
	% 		\toprule
	% 		\textbf{Approach} &Head-size & BLEU-4 & METEOR & ROUGE\_L & CIDER \\
	% 		\midrule
	% 		V-Transformer & 8& 42.55\% & 25.31\% & 53.14\% & 2.041 \\
	% 		NeuralCodeSum & 8& 43.42\% & 26.04\% & 54.05\% & 2.035 \\
	% 		MMTrans & 8& 45.10\% & 26.90\% & 55.21\% & 2.346 \\
	% 		SG-Trans & 8& 45.21\% & 27.11\% & 55.76\% & 2.301 \\
	% 		\hline
	% 		 & 2& 43.68\% & 27.81\% & 56.92\% & 2.457 \\
	% 		 & 4& 45.80\% & 28.01\% & 56.84\% & 2.512 \\
	% 		M2TS & 6& 46.01\% & 28.29\% & 57.13\% & 2.507 \\
	% 		 & 8& \textbf{46.83\%} & \textbf{28.93\%} & \textbf{57.87\%} & \textbf{2.573} \\
	% 	     & 10& 45.94\% & 27.86\% & 56.91\% & 2.462 \\
	% 		\bottomrule
	% 	\end{tabular}
	% 	}
	% \end{table}
	
	\section{Experimental Setup}
	
	\subsection{Datasets and Preprocessing}
	
	We conduct experiments on two Java and one Python datasets. For the two Java datasets, we call them JAH and JAL, and for the Python dataset we call it PYB. Specifically, we cite the 78422 pairs of $<$Java methods, comments$>$ used by Hu et al. \cite{hu2018summarizing} as the JAH dataset$\footnote{\url{https://github.com/xing-hu/TL-CodeSum}}$, which is collected in the repositories created on GitHub. The JAL dataset$\footnote{\url{http://leclair.tech/data/funcom/}}$ is a collection of 2.1 million pairs of java methods and summaries on the publicly available GitHub by LeClair et al. \cite{leclair2019recommendations}. 
	\begin{table}[htbp]
		\caption{Statistical analysis of the three datasets}
		\label{tab:table1}
		\resizebox{1\columnwidth}{!}{
			\begin{tabular}{lcccccccc}
				\toprule
				\multirow{2}{*}{Dataset} & \multicolumn{2}{c}{Code Length} & \multicolumn{2}{c}{NL Length}
				& \multicolumn{2}{c}{AST(Node)} & \multicolumn{2}{c}{AST(Depth)} \\
				\cmidrule(r){2-3} \cmidrule(r){4-5} \cmidrule(r){6-7} \cmidrule(r){8-9}
				& AvgC & UniC & AvgL &UniL & MaxN &AvgN & MaxD &AvgD  \\
				\midrule
				JAH & 120.99 & 23501 & 17.71 & 29645 & 2126 & 50 & 58 & 8.2 \\
				JAL & 65.23 & 207458 & 8.21 & 25341 & 1024 & 41 & 26 & 9.4 \\
				PYB & 57.12 & 125584 & 9.82 & 31116 & 1522 & 87 & 32 & 13.1 \\
				\bottomrule
			\end{tabular}
		}
	\end{table}
	Accordingly, we use the JDK$\footnote{\url{http://www.eclipse.org/jdt/}}$ compiler to parse the source code in the two Java datasets into ASTs \cite{hu2020deep}. The PYB dataset$\footnote{\url{https://github.com/EdinburghNLP}}$ we used is created by Barone et al. \cite{barone2017parallel}, and it contains 108726 pairs of python functions and their comments from the GitHub open source repositories. For Python functions, we use the Treelib$\footnote{\url{https://treelib.readthedocs.io/en/latest/}}$ toolkit to parse them as ASTs \cite{barone2017parallel}. In this paper, we use two Java datasets to explore the ability of M2TS to generate summaries for the same language source code of different lengths.

	Before the experiments, we preprocess the three datasets. We set the same length limits for code snippets and summaries as in \cite{hu2018summarizing, leclair2019recommendations, barone2017parallel}. The statistics for the three datasets are shown in Table \ref{tab:table1}. The parameters AvgC and AvgL in the table indicate the average length of code tokens and natural language, respectively, and the UniC and UniL indicate the number of unique tokens in the source code and natural language. In addition to code snippets and comments, we perform a statistical analysis of ASTs corresponding to the three datasets. The MaxN and AvgN denote the maximum and average number of nodes, and the MaxD and AvgD denote the maximum and average depth, respectively. To maintain the same segmentation settings as the compared baselines \cite{hu2018summarizing, leclair2019recommendations, barone2017parallel}, we split the PYB dataset into training, validation and testing sets in the proportion of 6 : 2 : 2, and the JAH and JAL datasets in the proportion of 8 : 1 : 1. To make the training and testing sets disjoint, we remove the duplicated samples from the testing set \cite{zhang2020retrieval}. In addition, we ensure that no identical samples exist for the two Java datasets. In common with \cite{ahmad2020transformer, zhang2020retrieval}, we split the source code tokens in the form of CamelCase and snake case in the three datasets into their respective subtokens. We show that this code tokens segmentation improves the quality of the generated summaries.
	
	In addition, we use a special symbol <PAD> to fill token sequences in the datasets with less than the maximum length and truncate sequences with greater than the maximum length. We set the maximum vocabulary size for source code and summaries to 50k, and tokens out of range are represented using the UNK \cite{zhang2020retrieval}. In particular, since the adjacency matrix cannot be obtained for ASTs of a single node, we remove ASTs with less than two nodes in the datasets. We also remove empty or one-word summaries because they do not fully describe the meaning of code snippets. 
	% During the model training, we add special markers <SOS> and <EOS> to denote the beginning and end of target sequences.
	% \vspace{-0.1cm}
	\subsection{Experimental Settings}
	
	We train M2TS using the SGD optimizer \cite{robbins1951stochastic} and set the initial learning rate to 0.0001. We set the minimum batch size and dropout to 32 and 0.2, respectively. We set the multi-head attention mechanism heads and the layers in both encoder and decoder to 8 and 6, $d\_model$ to 512, $d\_ff$ to 2048, $d\_v$ and $d\_k$ to 64. We train M2TS for at most 200 epochs and adopt early stop if the validation performance does not improve after 20 epochs. We leverage a beam search during generating summaries and set the beam size to 5. We set the initial feature vector dimension of the MSA module to 768  and the nonlinear activation function in GCN to the ReLu. The experiments in this paper are all conducted on the Intel(R) Xeon(R) Gold 5218 CPU @ 2.30GHz, 128GB RAM and TITAN RTX 24G GPU platform.
	
	\vspace{-0.1cm}
	\subsection{Evaluation Metrics}
	
	Similar to existing work \cite{hu2018deep, hu2020deep, zhang2020retrieval, wan2018improving, yang2021multi}, in this paper we evaluate the quality of the generated summaries using four evaluation metrics: BLEU \cite{papineni2002bleu}, ROUGE\_L \cite{lin2004rouge}, METEOR \cite{banerjee2005meteor} and CIDER \cite{vedantam2015cider}. They have commonly used metrics in tasks such as machine translation and text summarization.
	
	\textbf{BLEU} measures the average n-gram precision between the reference and generated summaries, with brevity penalty for short sentences \cite{wang2021cocosum}. In detail, the BLEU score is computed as:
	\begin{equation}
		BLEU-N = BP \cdot exp\sum\limits_{n=1}^N \omega_n\log p_n , 
	\end{equation}
	where $p_n$ is the score of $n-grams$ in the generated sentences which are present in the reference sentences, $\omega_n$ is the uniform weight $1/N$ and $BP$ is brevity penalty. In this paper, we set $N$ to 4, which is the maximum number of grams.
	
	\textbf{METEOR} is used to measure how the model captures the content from the reference sentences in the generated sentences \cite{wan2018improving}. Its calculation formula is as follows:
	\begin{equation}
		METEOR = (1- \gamma\cdot frag^\beta) \cdot \frac{P \cdot R}{\alpha \cdot P + (1-\alpha) \cdot R},
	\end{equation}
	where the $P$ and $R$ are the precision and recall, the $frag$ is a fragmentation fraction. The default values of the parameters $\gamma$, $\beta$, $\alpha$ are 0.5, 3.0 and 0.9, respectively.
	
	\textbf{ROUGE\_L} evaluates how much reference text appears in the generated text. Based on the longest common subsequence(LCS), it uses F-score which is the harmonic mean of precision and recall values \cite{wang2021cocosum}. Suppose $M$ and $N$ are generated and reference sentences of lengths $a$ and $b$, then:
	\begin{equation}
		P_L = \frac{LCS(M, N)}{a}, R_L = \frac{LCS(M, N)}{b}, 
	\end{equation}
	\begin{equation}
		F_{ROUGE\_L} = \frac{(1+\beta^2)P_L \cdot R_L}{R_L + \beta^2P_L},
	\end{equation}
	where the $\beta$ is set to 1.2 as in \cite{wan2018improving}, and the $F_{ROUGR\_L}$ is the value of ROUGE\_L.
	
	\textbf{CIDER} is a consensus-based evaluation metric for measuring image captioning quality \cite{zhang2020retrieval}. Then, the CIDER calculation formula is as follows:
	\begin{equation}
		CIDER(c, s) = \sum\limits_{n=1}^N w_n {\rm CIDER}_n(c, s),
	\end{equation}
	where the $c$ and $s$ are the generated and reference sentences, $n$ is set from 1 to 4 and the ${\rm CIDER}_n(c, s)$ score for $n-gram$ is computed using the average cosine similarity between $c$ and $s$.
	
	% In particular, the three evaluation metric scores BLEU, METEOR and ROUGE\_L are reported in percentages in the range [0, 1], while CIDER not in this range is usually reported as real values \cite{zhang2020retrieval}.
	Among them, BLEU and ROUGE\_L are based on precision and recall, respectively, while METEOR not only considers both, but also includes functions not found in other metrics, such as synonyms matching and homographs, etc. CIDER looks at whether the model captures key information.
	
	\begin{table*}
		%	\scriptsize
		%	\centering
		\caption{Comparison of our proposed approach with the baseline approaches}
		\label{tab:table2}
		% 	\resizebox{\textwidth}{25mm}{
		\resizebox{2.1\columnwidth}{!}{
			\begin{tabular}{lcccccccccccc}
				\toprule
				\multirow{2}{*}{\textbf{Approach}} & \multicolumn{4}{c}{JAH} & \multicolumn{4}{c}{JAL} & \multicolumn{4}{c}{PYB} \\
				\cmidrule(r){2-5} \cmidrule(r){6-9} \cmidrule(r){10-13}
				& BLEU-4 & METEOR & ROUGE\_L & CIDER & BLEU-4 & METEOR & ROUGE\_L & CIDER & BLEU-4 & METEOR & ROUGE\_L & CIDER  \\
				\midrule
				CODE-NN & 26.07\% & 13.97\% & 40.20\% & 0.855 & 16.25\% & 8.63\% & 38.24\% & 0.901 & 16.58\% & 9.08\% & 36.97\% & 0.988 \\
				Hybrid-DeepCom & 38.55\% & 23.28\% & 51.36\% & 1.254 & 20.74\% & 9.41\% & 40.31\% & 1.021 & 21.14\% & 9.64\% & 37.80\% & 1.284 \\
				Code2Seq & 38.86\% & 23.76\% & 51.85\% & 1.360 & 21.50\% & 10.23\% & 41.48\% & 1.175 & 21.86\% & 10.33\% & 38.13\% & 1.357 \\
				Code+GNN+GRU&  40.24\% & 24.25\% & 52.83\% & 1.855 & 22.13\% & 11.62\% & 43.32\% & 1.536 & 23.48\% & 12.31\% & 39.85\% & 1.766 \\
				V-Transformer &  42.55\% & 25.31\% & 53.14\% & 2.041 & 27.41\% & 16.47\% & 45.47\% & 2.068 & 30.58\% & 17.68\% & 43.04\% & 2.002 \\
				NeuralCodeSum&  43.42\% & 26.04\% & 54.05\% & 2.035 & 28.26\% & 18.25\% & 47.14\% & 2.185 & 31.07\% & 18.98\% & 45.50\% & 2.238 \\
				MMTrans&  45.10\% & 26.90\% & 55.21\% & 2.346 & 29.30\% & 18.06\% & 46.96\% & 2.267 & 32.36\% & 18.24\%  & 46.25\% & 2.316 \\
				SG-Trans&  45.21\% & 27.11\% & 55.76\% & 2.301 & 30.49\% & 18.81\% & 47.28\% & 2.362 & 32.21\% & 19.56\% & 46.34\% & 2.230 \\
				M2TS&  \textbf{46.84\%} & \textbf{28.93\%} & \textbf{57.87\%} & \textbf{2.573} & \textbf{31.63\%} & \textbf{20.06\%} & \textbf{49.31\%} & \textbf{2.504} & \textbf{33.84\%} & \textbf{21.83\%} & \textbf{47.92\%} & \textbf{2.411} \\
				
				\bottomrule
			\end{tabular}
		}
	\end{table*}
	
	% \vspace{-0.1cm}
	\subsection{Baselines}
	
	We compare M2TS with the existing source code summarization methods that are directly related to our work and described in detail below.
	
	\begin{itemize}
		\item \textbf{CODE-NN:} The method is first proposed by Iyer et al. \cite{iyer2016summarizing} using an end-to-end deep learning method. It is a typical encoder-decoder model that uses code tokens embedding as the output of the encoder and uses LSTM for the decoder combined with an attention mechanism to generate summaries.
		
		\item \textbf{Hybrid-DeepCom:} This method, proposed by Hu et al. \cite{hu2020deep}, uses the code token sequences and the AST sequences obtained via the SBT method as input to the Seq2Seq model, which can extract semantic and structural information of the code, respectively.
		
		\item \textbf{Code2Seq:} Alon et al. \cite{alon2018code2seq} used randomly paired paths of ASTs as model input and encoded each path as a fixed-length vector using LSTM. Finally, the summaries are generated at the decoder side with an attention mechanism.
		
		\item \textbf{Code+GNN+GRU:} This work is first used by LeClair et al. \cite{leclair2020improved} to use GNN for the code summarization task. They used GNN to embed ASTs and then input them into GRU in combination with the tokens features, after which an attention-based decoder generates summaries. 
		
		\item \textbf{V-Transformer:} To solve the long dependency problem that occurs when sequences are encoded, Vaswani et al. \cite{vaswani2017attention} proposed a Transformer model that entirely uses the self-attention mechanism, which can outperform previous models such as RNNs by using only the source code as input.
		
		\item \textbf{NeuralCodeSum:} Ahmad et al. \cite{ahmad2020transformer} used relative position encoding and copy mechanisms to improve the ordinary Transformer, which is the first approach to combine the Transformer with the code summarization task.
		
		\item \textbf{MMTrans:} Yang et al. \cite{yang2021multi} used two modalities of ASTs to generate code summaries. They used the AST sequences obtained via the SBT method and the  embedding representation obtained using GCN to extract the structure of ASTs.
		
		\item \textbf{SG-Trans:} Gao et al. \cite{gao2021code} injected local semantic information and global syntactic structure into the self-attentive module in the Transformer as an inductive bias to better capture the hierarchical features of source code.
	\end{itemize}
	
	% \vspace{-0.15cm}
	% We use these baselines for comparison with our work because CODE-NN is the first proposed method using neural networks. The baseline Hybrid-DeepCom traverses ASTs into sequences, Code2Seq uses the AST paths as input of the model, and Code+GNN+GRU uses GCN to embed ASTs. Hybrid-DeepCom, Code+GNN+GRU use the AST and source code modalities in a Seq2Seq model. In addition, V-Transformer and NeuralCodeSum are performed on the Transformer model. MMTrans and SG-Trans are studies of ASTs based on the Transformer. In contrast, our work is based on the Transformer using the AST and source code modalities to generate summaries, so it is appropriate to compare it to these baselines.
	
	% \vspace{-0.58cm}
	\section{Results and Analysis}
	
	Our study aims to assess whether M2TS outperforms the latest baselines and why the multi-scale and multi-modal method can improve the quality of the generated summaries. In this section, we give the experimental results and analysis of the proposed research problems.
	
	\begin{table*}
		%	\scriptsize
		%	\centering
		\caption{Comparison of setting different scales on the three datasets}
		\label{tab:table3}
		% 	\resizebox{\textwidth}{22mm}{
		\resizebox{2.1\columnwidth}{!}{
			\begin{tabular}{lccccccccccccc}
				\toprule
				\multirow{2}{*}{\textbf{Approach}} & \multirow{2}{*}{Scale} & \multicolumn{4}{c}{JAH} & \multicolumn{4}{c}{JAL} & \multicolumn{4}{c}{PYB} \\
				\cmidrule(r){3-6} \cmidrule(r){7-10} \cmidrule(r){11-14}
				& & BLEU-4 & METEOR & ROUGE\_L& CIDER & BLEU-4 & METEOR & ROUGE\_L& CIDER & BLEU-4 & METEOR & ROUGE\_L& CIDER  \\
				\midrule
				& 1  & 43.63\%  & 26.97\% & 55.39\%  & 2.100  & 29.15\% & 18.13\% & 47.20\% & 2.305 & 31.68\% & 19.23\% & 44.40\% & 2.184 \\
				& 2  & 44.82\% & 27.58\% & 57.03\% & 2.374 & 29.83\% & 18.74\% & 47.84\% & 2.376 & 32.82\% & 19.75\% & 45.48\% & 2.279 \\
				M2TS & 3 &  46.31\% & 28.10\% & 57.24\% & 2.501 & 30.57\% & 19.35\% & 48.46\% & 2.425 & 33.25\% & 20.37\% & 46.34\%  & 2.368 \\
				& 4&45.72\% & 27.21\% & 56.96\% & 2.485 & 29.64\% & 18.86\% & 47.83\% & 2.324 & 32.80\% & 19.87\% & 45.27\% & 2.231 \\
				& 5 &  45.08\% & 26.83\% & 55.23\% & 2.390 & 28.79\% & 18.17\% & 46.65\% & 2.255 & 31.75\% & 19.11\% & 45.63\% & 2.304 \\
				
				\hline
				M2TS (weights) & 3 & \textbf{46.84\%} & \textbf{28.93\%} &\textbf{57.87\%} & \textbf{2.573} & \textbf{31.63\%} & \textbf{20.06\%} & \textbf{49.31\%} & \textbf{2.504} & \textbf{33.84\%} & \textbf{21.83\%} & \textbf{47.92\%} & \textbf{2.411} \\
				M2TS w/o BERT  & 3 & 46.13\% & 28.52\% & 57.17\% & 2.519 & 30.47\% & 19.36\% & 48.50\% & 2.417 & 33.21\% & 20.75\% & 47.11\% & 2.335 \\
				M2TS (G-GCN)  & 3 & 45.96\% & 28.34\% & 57.03\% & 2.524 & 30.81\% & 19.57\% & 48.15\% & 2.446 & 33.05\% & 20.96\% & 46.81\% & 2.314 \\
				M2TS (attention)  & 3 & 45.75\% & 27.86\% & 56.58\% & 2.431 & 29.64\% & 18.62\% & 47.94\% & 2.386 & 32.47\% & 19.33\% & 46.24\% & 2.290 \\
				
				\bottomrule
				
			\end{tabular}
		}
	\end{table*}
	
	% \vspace{-0.1cm}
	\subsection{RQ1: How does our approach perform compared to the baselines?}
	We conduct experiments on two Java and one Python datasets, and the obtained experimental results are shown in Table \ref{tab:table2}. When reproducing baselines, we try our best to follow the data processing steps of baselines to prepare the input and set the hyper-parameters to be as consistent as possible with this paper. In addition, beam search is used for all baselines to ensure a fair comparison of experiments. 
	% Considering the uncertainty of the experimental results, we execute multiple tests on the datasets to prove the reliability of the model.
	
	As can be seen from the table, the metric scores obtained by M2TS on the three datasets are better than the latest baselines. For example, M2TS obtains a 46.84\% BLEU-4 score, 28.93\% METEOR score, 57.87\% ROUGE\_L score and 2.573 CIDER score on the JAH dataset. The METEOR score obtained by our model on the PYB dataset is 11.6\% higher than the baseline SG-Trans, which indicates that M2TS is better than the method used by SG-Trans to integrate code structure information in the Transformer. In addition, M2TS obtained higher metric scores than the baseline MMTrans on the three datasets. This is because MMTran does not use the source code as input, and although the reasons are given in the paper, the critical role of the source code still cannot be denied based on previous studies \cite{yang2021multi}. We use source code and ASTs as input and fuse the two modality features, thus achieving better metric scores than MMTrans. The BLEU-4 score obtained by M2TS on the JAH dataset is 3.42 higher than NeuralCodeSum. Furthermore, through experiments, we find that M2TS uses 1.6 times more parameters than NeuralCodSum, so this degree of improvement is worthwhile with a slight difference in parameters. V-Transformer and NeuralCodeSum yield higher scores than Code+GNN+GRU in the JAL dataset; the results confirm that the Transformer can handle long sequence input problems well. The score of Code+GNN+GRU is improved compared to Code2Seq and Hybrid-DeepCom, and on the JAL dataset Code+GNN+GRU is 13.6\% higher than Code2Seq on the METEOR metric and 7.5\% higher than Hybrid-DeepCom on the ROUGE\_L metric. It can be seen that embedding the AST in the form of a graph is better than using the AST sequences and AST paths, as the AST graph contains more structural information. In addition, we also compare with the earliest deep learning method CODE-NN, and it is not hard to find that inputting only source code into the Seq2Seq model yields much lower scores than Hybrid-DeepCom. This also confirms that the structural features of ASTs are significant for the source code summarization task. As can be seen from the data throughout Table \ref{tab:table2}, the quality of the generated summaries is directly related to the completeness and accuracy of the code structure extraction and the comprehensiveness of the code modality features. Furthermore, Table \ref{tab:table2} shows that the metric scores obtained on the PYB and JAL datasets are generally lower than those obtained on the JAH dataset, which may be related to the source code themselves or ASTs parsed by them.
	
	% \vspace{-0.25cm}
	\subsection{RQ2: How does the number of scales in the MSA module affect the performance of M2TS?}
	
	In the multi-scale AST feature extraction method, we use the power matrices of the adjacency matrix to represent the different scales of ASTs. This section explores the effect of different scales on the experimental results on the three datasets.
	
	By conducting experiments on the three datasets, we find that the highest metric scores are obtained when the number of scales is 3. As can be seen from Table \ref{tab:table3}, the BLEU-4 score obtained with the number of scales set to 3 is 6.1\% higher than that obtained with a scale of 1 on the JAH dataset. It can be seen that using three scales provides a more comprehensive representation of the structural features of source code, making the generated summaries more fluent in syntax. In the other three evaluation metrics of the corresponding datasets, the scores of scale 3 are also the highest. As the number of scales increases, more structural features can be extracted from ASTs. When the number of scales is set to 4 or 5, the scores of the four evaluation metrics are decreased. We consider that when too many scales are set, some noisy data will be included (e.g., some duplicate paths information will be introduced), resulting in poor summaries obtained at decoding. 
	
	\begin{table}[htbp]
		\caption{Comparison of setting different heads on the JAH dataset}
		\label{tab:table4}
		\resizebox{1\columnwidth}{!}{
			\begin{tabular}{lccccc}
				\toprule
				\textbf{Approach} &Head-size & BLEU-4 & METEOR & ROUGE\_L & CIDER \\
				\midrule
				V-Transformer & 8& 42.55\% & 25.31\% & 53.14\% & 2.041 \\
				NeuralCodeSum & 8& 43.42\% & 26.04\% & 54.05\% & 2.035 \\
				MMTrans & 8& 45.10\% & 26.90\% & 55.21\% & 2.346 \\
				SG-Trans & 8& 45.21\% & 27.11\% & 55.76\% & 2.301 \\
				\hline
				& 2& 43.68\% & 27.81\% & 56.92\% & 2.457 \\
				& 4& 45.80\% & 28.01\% & 56.84\% & 2.512 \\
				M2TS & 6& 46.01\% & 28.29\% & 57.13\% & 2.507 \\
				& 8& \textbf{46.83\%} & \textbf{28.93\%} & \textbf{57.87\%} & \textbf{2.573} \\
				& 10& 45.94\% & 27.86\% & 56.91\% & 2.462 \\
				\bottomrule
			\end{tabular}
		}
	\end{table}
	
	In addition, instead of directly summing the outputs of each scale in the multi-scale AST feature extraction module, we assign different weights to them. The reason for doing this is that we use the output of the previous scale as input to the following scale, and simply adding up the output of each scale would contain some duplicate information. From Table \ref{tab:table3}, it is seen that the metric scores obtained by using the weighting method when setting the same scale are higher. Taking the PYB dataset as an example, the weighted metric scores obtained are 3.4\% higher on ROUGE\_L than the unweighted M2TS. It is experimentally verified that the best quality of the generated summaries is obtained when the output weights of the three scales are set to $0.1, 0.2, 0.7$ respectively. Since we use the output of the previous scale as the input of the next scale, it makes sense to set the output weights for the third scale. In particular, it can be seen from the experimental results that the best results are obtained by the multi-scale method for different code length datasets, so it is known that our method has strong universality.
	
	% \vspace{-0.25cm}
	\subsection{RQ3: How do the multi-head attention mechanism heads affect the performance of M2TS?}
	
	\begin{table}[htbp]
		\centering
		\caption{Qualitative example of the different models' performance on the three datasets}
		% \resizebox{\textwidth}{2mm}{
		% \setlength{\tabcolsep}{0.65mm}{
		\resizebox{1.0\columnwidth}{!}{
			\begin{tabular}{c|l}
				% \small
				\toprule
				Dataset & Example   \\
				\midrule
				& protected void addtogui(jpanel gui, jtextfield b, string cmd) \{  \\
				& \quad b.setactioncommand(cmd); \\
				& \quad b.addactionlistener(this); \\
				& \quad gui.add(b); \\
				& \} \\
				JAH & NeuralCodeSum: adds the feature to the attribute object. \\
				& M2TS (1): adds a attribute of the object. \\
				& M2TS (attention): adds a feature to the attribute. \\
				& M2TS: adds a feature to the gui attribute of the object. \\
				& Reference: adds a feature to the gui attribute of the layer object. \\
				\midrule
				& def traverse\_tree(course):  \\
				& \quad queue = [course] \\
				& \quad while(len(queue)>0): \\
				& \qquad node = queue.pop() \\
				& \qquad queue.extend(node.get\_children()) \\
				& \quad reture True \\
				PYB & NeuralCodeSum: load the descriptor from the descriptor. \\
				& M2TS (1): load the node descriptor. \\
				& M2TS (attention): load a descriptor to the course. \\
				& M2TS: load every descriptor in the course. \\
				& Reference: load every descriptor in course. \\
				\midrule
				& public int getSquareWidth() \{  \\
				& \quad if(skin == null) \{ \\
				& \qquad return 0; \\
				& \quad \} \\
				& \quad return skin.graphicWidth; \\
				& \} \\
				JAL & NeuralCodeSum: returns the current skin of the current called width. \\
				& M2TS (1): returns the current skin of square. \\
				& M2TS (attention): returns the width of the scaled skin. \\
				& M2TS: returns the current width of square from scaled skin. \\
				& Reference: returns the current width of square from the scaled skin. \\
				\bottomrule
			\end{tabular}%
		}
		\label{tab:table5}%
	\end{table}%
	
	M2TS uses the multi-head attention mechanism and the fully connected network in the Transformer as the basic structure of the encoder and decoder. Using multiple heads in the multi-head attention mechanism allows extra attention to be applied to different parts of the input information. However, too many or too few heads can impact the experimental results. To explore the optimal number of heads, we set the number of heads to 2, 4, 6, 8, and 10 for experiments. The results find that the highest metric scores are achieved for the three datasets when the number of heads is 8. For convenience, in Table \ref{tab:table4} we only list the experimental results obtained by setting different head numbers on the JAH dataset. As shown from the table, M2TS is 6.7\% higher than the baseline SG-Trans on METEOR when the same number of heads is set. The score increases with the number of heads until the number of heads is set to 8, after which it starts to decrease so that more heads in the multi-head attention is not better. We consider that when too many heads are set, it leads to distraction and thus destroys the correlation between the data. The other baselines are not listed in Table \ref{tab:table4} because they do not use the multi-head attention mechanism.
	
	\subsection{Ablation Study}
	We further conduct ablation experiments to verify the effects of the multi-scale feature extraction method and the cross modality fusion method on M2TS. The results are shown in the bottom half of Table \ref{tab:table3}.
	
	\textbf{Analysis of using the BERT pre-training.} In the multi-scale AST feature extraction module, we embed the "\textit{type}" and "\textit{value}" of nodes using the BERT pre-training method, and input them into GCN as the initial feature vector of nodes. In this section, we compare it with the random initialization method for nodes. As seen in Table \ref{tab:table3}, higher metric scores are obtained using BERT than the random initialization method on the three datasets. This is since that AST nodes contain code tokens information, and embedding them as the initial feature vector of nodes can preserve code tokens semantic features as much as possible.
	
	\textbf{Analysis of GCN with the residual connection.} To extract the structural features of ASTs more completely and accurately, we improve the general GCN in the multi-scale method by using a residual connection to sum the output of the previous graph convolution layer with the production of the current layer. In order to verify the effectiveness of this method, we use residual-free GCN with it for ablation analysis. The M2TS (G-GCN) represents using the general GCN in Table \ref{tab:table3}, and the results show that the GCN with residual connection works better. This may be because the general GCN computes fewer message-passing iterations when embedding, resulting in some loss of features \cite{peng2021integrating}.
	
	\textbf{Analysis of the cross modality fusion.} To confirm that the ACF module is better than using the traditional fusion method, we fuse AST and code tokens features using two traditional attention mechanisms as in \cite{hu2020deep, leclair2020improved}. Then the fused features are fed into the decoder to generate summaries, where we use the GRU for our decoder. The ablation comparison experiments find that the metric scores obtained using the ACF module are higher on the three datasets. As shown in Table \ref{tab:table3}, the M2TS (attention) denotes the summary generated using the traditional fusion method, and the ACF module yields 4.9\% higher CIDER scores on the JAL dataset than using the traditional method. This shows that our cross modality fusion method can highlight the critical information in each modality to help generate summaries.
	
	% \vspace{-0.8cm}
	\subsection{Qualitative Analysis}
	Since the statistics do not fully reveal the predictive power of the model, we further discuss the quality of the summaries generated by M2TS through qualitative analysis. Table \ref{tab:table5} shows the summaries generated by the different models for the examples in the three datasets, where the model M2TS (1) denotes the summaries generated when the scale is set to 1. We chose the NeuralCodeSum for comparison because this baseline played a key transition role throughout the study, with subsequent studies MMTrans and SG-Trans being based on the Transformer. From Table \ref{tab:table5}, we can observe that M2TS can generate better descriptions for the given programs. Specifically, for the example in the JAH dataset, the keyword "\textit{gui}" is missing in the results of the baseline NeuralCodeSum and ablation analysis but can be generated accurately by our model. In addition, for the examples in the PYB and JAL datasets, our model can generate semantically coherent phrases "\textit{every descriptor}" and "\textit{current width}", which is not possible with other listed baselines. M2TS generates high-quality summaries for the following reasons: first, embedding ASTs at multiple scales enables more comprehensive extraction of syntactic structural information of ASTs, then the fusion of two modalities, ASTs and code tokens, using the ACF module can learn the contextual semantic relevance of code tokens while highlighting structural features of source code. Due to the two main components, the generated summaries are semantically more fluently compared to NeuralCodeSum and other baselines as seen in the examples.  
	
	\section{Related Work}
	
	\subsection{Source Code Summarization}
	
	Source code summarization aims to generate short summaries for code snippets. Existing code summarization approaches can be divided into traditional and based on deep learning.
	
	% \begin{figure}[htbp]
	% 	\centering
	% 	\includegraphics[width=0.9\linewidth]{Figure_1.png}
	% 	\caption{Human evaluation results, where “CGG” and “NCS” indicate the baseline Code+GRU+GNN and NeuralCodeSum. }
	% 	\label{fig:figure5}
	% \end{figure}
	
	Among the traditional methods of source code summarization, the earliest is the approach based on artificial templates. Sridhara et al. \cite{sridhara2010towards} used the SWUM to create a rule-based approach to generating summaries for Java methods. Haiduc et al. \cite{haiduc2010supporting, haiduc2010use} applied text retrieval techniques and latent semantic search to select essential keywords from source code and treated these keywords as comments. Hindle et al. \cite{hindle2016naturalness} addressed the naturalness of code languages and proved that source code could be modelled by probabilistic models. Subsequently, Dana et al. \cite{movshovitz2013natural} combined statistical probability models such as topic models and n-gram models to predict comments for Java methods \cite{hu2020deep}.
	
	Since deep learning can automatically learn pattern features from large-scale data, some studies have experimented with deep learning-based approaches to generate code summaries \cite{wan2018improving, shido2019automatic, mou2016convolutional, mcburney2014automatic}. Inspired by Neural Machine Translation (NMT), some studies have looked at code summarization as a variant of the translation task. Lyer et al. \cite{iyer2016summarizing} proposed a Seq2Seq model based on attention mechanism, where they used token embedding vectors as the output of the encoder and then used the LSTM as the decoder to generate summaries. With the advent of code-structured representation ASTs, some research has begun to revolve around them, with Hu et al. \cite{hu2018deep, hu2018summarizing}, Zhang et al. \cite{zhang2019novel}, and LeClair et al. \cite{leclair2019neural} successively using ASTs as input to the encoder-decoder model. In addition, Wan et al. \cite{wan2018improving} used deep reinforcement learning to enhance code summarization as 
	the way to address exposure bias during coding \cite{wang2020reinforcement}. LeClair et al. \cite{leclair2020improved} used both AST and source code to generate comments for Java methods, which differ from their work published in 2019 ICSE \cite{leclair2019neural} in that they studied AST graphs rather than AST sequences. With the emergence of the Transformer model \cite{vaswani2017attention}, Ahmad et al. \cite{ahmad2020transformer} improved it to generate summaries for both Java and Python language. Due to the power of the Transformer model, much work has subsequently emerged around it \cite{wang2021cocosum, gao2021code, zugner2021language, feng2020codebert}.
	
	Compared with the above works, our approach can comprehensively extract the structural features of source code and fuse code token and AST features in a cross modality fusion method to obtain the best performance than the above state-of-the-art methods.
	
	% \vspace{-0.25cm}
	\subsection{Graph Convolutional Neural Network}
	GCN is a graph embedding algorithm used more often in recent years. Its core idea is the weighted summation of the feature values of nodes themselves and their neighboring nodes \cite{kipf2016semi}, where the feature value refers to the content information of nodes. Each layer of graph convolution has two main inputs: the output of the previous layer of graph convolution and the adjacency matrix. Since there is no previous layer output when inputting the first layer of graph convolution, the input is the initialization feature vector of nodes. Before inputting the GCN, the adjacency matrix needs to be normalized, and the following is the formula:
	\begin{equation}
		\hat{A} = \tilde{D}^{-1/2}\tilde{A}\tilde{D}^{-1/2},
	\end{equation}
	where $ \tilde{A}=A + I $, $I$ denotes the unit matrix, which is added to consider the information of the nodes themselves. $\tilde{D}$ is the degree matrix of $\tilde{A}$, and $\hat{A}$ is the normalized adjacency matrix.
	
	In this paper, we use the BERT pre-training \cite{devlin2018bert} to embed nodes and input the embedding vectors as the initial feature vectors of nodes along with the normalized adjacency matrix into the GCN. In addition, we add a residual connection to the general GCN to prevent features loss when performing the next graph convolution layer.
	
	\section{Threats To Validity}
	
	We have identified the following threats to validity:
	\begin{itemize}
		\item {\textbf{Evaluation method.}} In this paper, we use four popular automatic evaluation metrics to assess the similarity of generated and reference sentences. Although they are computationally fast and widely used in this field \cite{leclair2020improved}, they have some limitations, such as the valid summary generated by the model may not be aligned with the reference summary or the generated summary is aligned with the reference summary but not readable. Therefore, it is necessary to evaluate generated summaries from additional perspectives, such as human evaluation \cite{hu2020deep, wan2018improving}.
		\item {\textbf{Hardware limitations.}} Our experiments cannot perform hyper-parameter optimizations of the model due to hardware limitations \cite{leclair2020improved}. In Table \ref{tab:table2}, our model or baselines may be affected by certain hyper-parameters (e.g., learning rate and batch size), caused different metric scores or rankings. This is an unavoidable problem in deep learning research. The hyper-parameters we set when reproducing the work in baselines should match their descriptions as much as possible to mitigate this problem \cite{wan2018improving}.
		\item {\textbf{Dataset type.}} We use the Java and Python datasets to validate the feasibility of the model in this paper. Although our proposed multi-scale AST feature extraction method is set at the same scales in these three datasets, we do not know whether this setup is equally applicable to other datasets. Therefore, it is necessary to conduct experiments on more large datasets (e.g., C, C\#) to verify the reliability of this method. 
	\end{itemize}
	
	\section{Conclusion}
	In this paper, we propose a multi-scale multi-modal approach based on the Transformer for source code summarization (M2TS), which present a multi-scale AST feature extraction method to extract the structural features of source code more completely and accurately at multiple local and global levels. In addition, we use a new cross modality feature fusion method to fuse the encoded AST features with the code tokens features, highlighting the structural information of source code and learning the contextual semantic relevance between code tokens. Finally, we input the fused features and the encoded code tokens features into the combined decoder to generate summaries. In particular, our extensive experiments on two Java and one Python datasets demonstrate the effectiveness of M2TS, and the results show that it outperforms the related approaches.
	
	%%
	%% The acknowledgments section is defined using the "acks" environment
	%% (and NOT an unnumbered section). This ensures the proper
	%% identification of the section in the article metadata, and the
	%% consistent spelling of the heading.
	\begin{acks}
		This work is financially supported by the Natural Science Foundation of Shandong Province, China (ZR2021MF059, ZR2019MF071), National Natural Science Foundation of China (61602286, 61976127) and Special Project on Innovative Methods (2020IM020100).
	\end{acks}
	
	%%
	%% The next two lines define the bibliography style to be used, and
	%% the bibliography file.
	% \bibliographystyle{ACM-Reference-Format}
	\balance
	\bibliography{icpc}
	
	%%
	%% If your work has an appendix, this is the place to put it.
	
\end{document}